\documentclass[fleqn,11pt,twoside]{article}

\usepackage{amsthm,amsthm,amssymb, color, xcolor,epsfig, graphics, subfigure}
\usepackage[normalem]{ulem}
\usepackage{arydshln, mathrsfs}
\usepackage{amsmath, graphicx, latexsym, lscape, wrapfig}

\makeatletter
\newcommand{\copyrightnote}[2]{{\renewcommand{\thefootnote}{}
 \footnotetext{\small\it
\begin{flushleft}
 \copyright \ #1   #2  
\end{flushleft}}}}

\newcommand{\Name}[1]{\begin{flushleft}
                       \LARGE \bf #1
                       \end{flushleft}\vspace{-3mm}}

\newcommand{\Author}[1]{\begin{flushleft}
                       \it #1 \end{flushleft}}

\newcommand{\Address}[1]{\begin{flushleft}
                       \it #1 \end{flushleft}}

\newcommand{\Date}[1]{\begin{flushleft}
                      \small  \it #1 \end{flushleft}}

\newcommand{\evenhead}{Author \ name}
\newcommand{\oddhead}{Article \ name}

\renewcommand{\@evenhead}{
\hspace*{-3pt}\raisebox{-15pt}[\headheight][0pt]{\vbox{\hbox to \textwidth
{\thepage \hfil \evenhead}\vskip4pt \hrule}}}
\renewcommand{\@oddhead}{
\hspace*{-3pt}\raisebox{-15pt}[\headheight][0pt]{\vbox{\hbox to \textwidth
{\oddhead \hfil \thepage}\vskip4pt\hrule}}}
\renewcommand{\@evenfoot}{}
\renewcommand{\@oddfoot}{}

\long\def\@makecaption#1#2{%
  \vskip\abovecaptionskip
  \sbox\@tempboxa{\small \textbf{#1.}\ \ #2}%
  \ifdim \wd\@tempboxa >\hsize
    {\small \textbf{#1.}\ \ #2}\par
  \else
    \global \@minipagefalse
    \hb@xt@\hsize{\hfil\box\@tempboxa\hfil}%
  \fi
  \vskip\belowcaptionskip}

\newcommand{\JNMPnumberwithin}[3][\arabic]{%
  \@ifundefined{c@#2}{\@nocounterr{#2}}{%
    \@ifundefined{c@#3}{\@nocnterr{#3}}{%
      \@addtoreset{#2}{#3}%
      \@xp\xdef\csname the#2\endcsname{%
        \@xp\@nx\csname the#3\endcsname .\@nx#1{#2}}}}%
}

\newcommand{\resetfootnoterule} {
  \renewcommand\footnoterule{%
  \kern-3\p@
  \hrule\@width.4\columnwidth
  \kern2.6\p@}
}

\renewcommand{\footnoterule}{}

\makeatother

\theoremstyle{definition}

\begin{document}

\renewcommand{\evenhead}{ {\LARGE\textcolor{blue!10!black!40!green}{{\sf \ \ \ ]ocnmp[}}}\strut\hfill 
M Hamanaka and S-C Huang
}
\renewcommand{\oddhead}{ {\LARGE\textcolor{blue!10!black!40!green}{{\sf ]ocnmp[}}}\ \ \ \ \  
Solitons in 4d Wess-Zumino-Witten models
}

\thispagestyle{empty}
\newcommand{\FistPageHead}[3]{
\begin{flushleft}
\raisebox{8mm}[0pt][0pt]
{\footnotesize \sf
\parbox{150mm}{{Open Communications in Nonlinear Mathematical Physics}\ \ \ \ {\LARGE\textcolor{blue!10!black!40!green}{]ocnmp[}}
\quad Special Issue 2, 2024\ \  pp
#2\hfill {\sc #3}}}\vspace{-13mm}
\end{flushleft}}

\FistPageHead{1}{\pageref{firstpage}--\pageref{lastpage}}{ \ \ }

\strut\hfill

\strut\hfill

\copyrightnote{The author(s). Distributed under a Creative Commons Attribution 4.0 International License}

\begin{center}

{\bf {\large Proceedings of the OCNMP-2024 Conference:\\ 

\smallskip

Bad Ems, 23-29 June 2024}}
\end{center}

\smallskip

\Name{Solitons in 4d Wess-Zumino-Witten models \\-- Towards unification of integrable systems --}

\Author{Masashi Hamanaka and Shan-Chi Huang\footnote{First author}}

\Address{
Graduate School of Mathematics, Nagoya University, Nagoya, 464-8602, JAPAN
}

\Date{Received August 30, 2024; Accepted September 16, 2024}

\setcounter{equation}{0}

\begin{abstract}

\noindent 
We construct soliton solutions of the four-dimensional Wess-Zumino-Witten (4dWZW) model in the context of a unified theory of integrable systems with relation to the 4d/6d Chern-Simons theory. We calculate the action density of the solutions and find that the soliton solutions behave as the KP-type solitons, that is, the one-soliton solution has a localized action/energy density on a 3d hyperplane in 4-dimensions (soliton wall) and the $n$-soliton solution describes $n$ intersecting soliton walls with phase shifts. We note that the Ward conjecture holds mostly in the split signature $(+,+,-,-)$. Furthermore, the 4dWZW model describes the string field theory action of the open N=2 string theory in the four-dimensional space-time with the split signature and hence our soliton solutions would describe a new-type of physical objects in the N=2 string theory. We discuss instanton solutions in the 4dWZW model as well. Noncommutative extension and quantization of the unified theory of integrable systems are also discussed.

\end{abstract}

\label{firstpage}


\section{Introduction}

Four-dimensional Wess-Zumino-Witten 
models are 
analogues of the two-dimensional WZW models and possess aspects of conformal field theory \cite{IKUX, LMNS, NaSc, NaSc2, Nekrasov}. 
Equation of motion of the 4dWZW 
model is the Yang equation which is equivalent   
to the anti-self-dual Yang-Mills (ASDYM) equation \cite{Donaldson}.  
It is well known for the Ward conjecture that 
the ASDYM equations can be reduced to many 
classical integrable systems, such as the Korteweg–de Vries (KdV) equation,  
Toda equation and Painlev\'e equations. \cite{Ward, MaWo}. 
Furthermore, integrability of the ASDYM equations can be understood in the framework of twistor theory and hence 4d WZW model possesses aspects of the twistor theory as well.

\medskip

\begin{wrapfigure}[11]{r}{8cm}
\centering
\includegraphics[width=8.4cm]{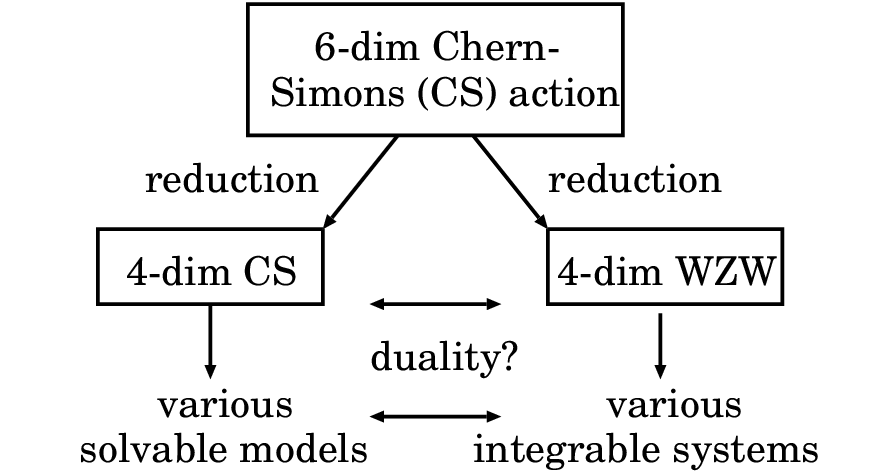}
\label{unify}
\end{wrapfigure}

On the other hand, 4d Chern-Simons (CS) theory has connections to many solvable models such as spin chains and principal chiral models e.g. \cite{CoYa, DLMV, FSY}. These two theories (4dCS and 4dWZW) can be derived from a 6dCS theory like a ``double fibrations'' as in the above figure \cite{Costello, BiSk}. This fact suggests a nontrivial duality correspondence between the 4dWZW model and the 4dCS theory. We note that the Ward conjecture holds mostly for the split signature $(+,+,-,-)$  which corresponds to the four-dimensional space-time for open N=2 string theory \cite{OoVa, Marcus, Berkovits}. Hence a unified theory of integrable systems (6dCS$\rightarrow$4dCS/4dWZW) can be proposed in the context of the split signature\footnote{This case of (6dCS$\rightarrow$4dCS/4dWZW) is also discussed in \cite{BiSk, Bittleston}.} and classical solutions of 4dWZW model describe classical physical objects in the open N=2 string theory. 

In this paper, we discuss several classical solutions of the ASDYM equations in the context of 4dWZW models, rather than the Yang-Mills theory. In the split signature, all of them can be interpreted as classical physical objects in open N=2 string theory. Firstly, we review our previous work on soliton solutions in the 4dWZW model. The action densities of the soliton solutions show that the behaviors of the soliton solutions are quite similar to the Kadomtsev-Petviashvili 
(KP) solitons, more explicitly, one-soliton solutions are codimension-one solitons whose action densities are localized on three-dimensional hyperplanes in four dimensions. $n$-soliton solutions can be interpreted as a ``non-linear superposition'' of $n$ one-soliton solutions with phase shifts.
Secondly, we discuss instanton solutions in the 4dWZW model for the four-dimensional Euclidean signature.  

This paper is organized as follows. In section 2, the 4dWZW model is introduced and our conventions are set up. In section 3, soliton solutions of the Yang equation are introduced by the method of Darboux transformations. In section 4, the behaviors of the soliton solutions 
are discussed in the sense of action density. These three sections can be considered as a brief review of the paper \cite{HHK}. 
In section 5, we discuss the equivalence between ASDYM equation and Yang equation and then derive $G=SL(2, \mathbb{C})$ instanton solutions of the Yang equation from $G_{\scriptsize{\mbox{YM}}}=SU(2)$ Yang-Mills instantons. These are new results in this paper. 
In section 6, some examples of the Ward conjecture are reviewed.  Section 7 is devoted to conclusion and discussion where new study directions towards unification of integrable systems are proposed.

\section{Four-Dimensional Wess-Zumino-Witten Model}


In this section, we review the four-dimensional Wess-Zumino-Witten (4dWZW) model.
In order to discuss various signatures in a unified way, we introduce a four-dimensional space $M_4$ with complex coordinates $(z,\widetilde z, w, \widetilde w)$ and the flat metric:
\begin{eqnarray}
ds^2&=&g_{mn}dz^{m}dz^{n}=2(dz d\widetilde z -dw d\widetilde w),
~~~m,n\in \{1,2,3,4\}
\label{com_metric}
\end{eqnarray}
where $(z^{1},z^{2},z^{3},z^{4}):=(z,\widetilde z, w, \widetilde w)$, 
that is, $g_{12}=g_{21}=-g_{34}=-g_{43}=1$ and $g_{mn}=0$ otherwise. 
The space $M_4$ can be reduced to three kinds of real spaces by imposing suitable reality conditions (real slices) on $(z,\widetilde z, w, \widetilde w)$. For example, the Euclidean real space $\mathbb{E}$ ($\mathbb{R}^4$ with the Euclidean signature) is given by the real slice $\widetilde z=\overline z,~\widetilde w= -\overline w$. 
The Ultrahyperbolic real space $\mathbb{U}$ ($\mathbb{R}^4$ with the split signature) is given by the real slices: $z, \widetilde z,w, \widetilde w\in \mathbb{R}$ or $\widetilde z=\overline z, ~\widetilde w= \overline w$, which are denoted respectively by $\mathbb{U}$ and $\mathbb{U}^\prime$. In this paper we do not consider the case of $\mathbb{U}^\prime$ because unitarity condition of $J$ leads to trivial action densities \cite{HHK}. Our choices of the real slices are as follows:
\begin{eqnarray}
&\mathbb{E}:& 
\sqrt{2}(z,\widetilde z, w, \widetilde w)=(
x^{1}+ix^{2},
x^{1}-ix^{2}, 
x^{3}+ix^{4}, 
-(x^{3}-ix^{4})),
\label{E}
\label{Reality condition_E}
\\
\label{U}
\label{Reality condition_U}
&\mathbb{U}:& 
\sqrt{2}(z,\widetilde z, w, \widetilde w)=(
x^{1}+x^{3}, 
x^{1}-x^{3}, 
x^{4}+x^{2}, 
x^{4}-x^{2}).
\end{eqnarray}
In this paper, we mainly consider the Ultrahyperbolic space $\mathbb{U}$ in sections 2--4 and the Euclidean space $\mathbb{E}$ in section 5.


The action of the 4dWZW model consists of 
two parts and given as follows:
\begin{eqnarray}
\label{WZW4}
S_{\scriptsize {\mbox{4dWZW}}}&:=&
S_{\scriptsize{\sigma}}+S_{\scriptsize {\mbox{WZ}}},
\\
S_{\scriptsize{\sigma}}&:=&
\frac{i}{4\pi}\int_{M_4}\omega \wedge 
\mathrm{Tr}\left[
\left(\partial J\right)J^{-1} \wedge (\widetilde{\partial} J)J^{-1}
\right], \\
\label{WZ}
S_{\scriptsize {\mbox{WZ}}}
&:=&
-
\frac{i}{12\pi}\int_{M_4} A \wedge
\mathrm{Tr}\left[
\left(dJ\right)J^{-1}\wedge\left(dJ\right)J^{-1}\wedge\left(dJ\right)J^{-1} 
\right], 
\end{eqnarray}
where the dynamical variable $J$ is a smooth map from $M_4$ to 
$G=GL(N,\mathbb{C})$,  
and $\omega$ is the K\"ahler two-form on $M_4$ given by
\begin{eqnarray}
\label{kahler}
\omega =\frac{i}{2}\left( dz \wedge d\widetilde{z} - dw \wedge d\widetilde{w}\right),
\end{eqnarray}
and the one-form is chosen as $A=(i/4)\left(zd\widetilde{z}-wd\widetilde{w}\right)$ so that $\omega=dA$. 
The exterior derivatives are defined as follows:
\begin{eqnarray}
d:=\partial + \widetilde{\partial},~~~
\partial:=dw\partial_w + dz\partial_z,~~~
\widetilde{\partial}:=d\widetilde{w}\partial_{\widetilde{w}} + d\widetilde{z}\partial_{\widetilde{z}}.
\end{eqnarray}
The first part $S_{\scriptsize{\sigma}}$ is called the non-linear sigma model (NL$\sigma$M) term or simply the sigma model term, 
and the second part 
is called the Wess-Zumino (WZ) term.

The equation of motion of the 4dWZW model \eqref{WZW4} is the Yang equation:
\begin{eqnarray}
\label{Yang}
\widetilde{\partial}\left(
\omega \wedge \left(\partial J \right)J^{-1} \right)=0
~~~\Longleftrightarrow~~~
\label{yang}
\partial_{\widetilde{z}}((\partial_z J) J^{-1})
-\partial_{\widetilde{w}}((\partial_w J) J^{-1} )=0. 
\end{eqnarray} 
This is equivalent to the anti-self dual Yang-Mills equation as discussed in section 5.1. 
We note that the following transformation acts on the Yang equation covariantly: 
\begin{eqnarray}
\label{PQ}
J \mapsto PJQ, 
\end{eqnarray} 
where $P(z,w)$ and $Q(\widetilde{z},\widetilde{w})$ be a $G$-valued function. 

The 4dWZW action density (\ref{WZW4}) can be explicitly written 
in the flat four-dimensional real spaces as follows 
\begin{eqnarray}
S_{\sigma}
&=&-\frac{1}{16\pi}
\int_{M_4}
\mbox{Tr}\left[
\left( \partial_m J \right)J^{-1}
\left( \partial^{m} J\right)J^{-1}\right]
dz \wedge d\widetilde{z} \wedge dw \wedge d\widetilde{w}, 
\label{Tr(A_m A_n)_differential form}
\\
&=& 
\label{S_sigma}
\int_{\mathbb{U}\scriptsize{\mbox{or}}\mathbb{E}}
{\cal {L}}_\sigma
dx^{1}\wedge dx^{2}\wedge dx^{3}\wedge dx^{4},\\
\label{L_sigma}
{\cal{L}}_\sigma&:=&-\frac{1}{16\pi}\mbox{Tr}\left[
\left( \partial_{\mu} J \right)J^{-1}
\left( \partial^{\mu} J \right)J^{-1}\right]. 
\end{eqnarray}
where $\partial^m := g^{mn}\partial_n$ and the metrics are given by 
 (\ref{com_metric}). 
The NL$\sigma$M action density is denoted by ${\cal{L}}_\sigma$.
Similarly the Wess-Zumino action density ${\cal{L}}_{\scriptsize{\mbox{WZ}}}$ can be read from the Wess-Zumino action \eqref{WZ}. (cf. \eqref{TrA^3}). 
The NL$\sigma$M action density is positive definite in the case of  $G=U(N)$ and $\mathbb{E}$.

The 4dWZW model has conformal field theory aspects such as 
the Polyakov-Wiegmann formula \cite{NaSc, NaSc2}  
and the Sugawara-like construction of current algebras \cite{IKUX, NaSc, NaSc2}.

\section{Darboux Transformation and Soliton Solutions}

In this section, we review the soliton solutions of 
the Yang equation  
which are constructed 
by applying the Darboux transformation \cite{GNO, GHHN}. 

\subsection{Darboux Transformation for Yang Equation}

In this subsection, we focus on $G=GL(N,\mathbb{C})$. 
A Lax representation of (\ref{yang}) is 
given by the following linear system \cite{GNO}:
\begin{eqnarray}
L(f)&:=&
J \partial_{w}(J^{-1} f)
- (\partial_{\widetilde{z}}f)\zeta=0,\nonumber\\
M(f)&:=&
J \partial_{z}(J^{-1} f)
- (\partial_{\widetilde{w}}f)\zeta=0. 
\label{lin_yang}
\end{eqnarray}
The spectral parameter $\zeta$ here has been generalized to 
an $N\times N$ constant matrix so that the Darboux transformation can be a non-trivial transformation. 
The compatibility condition $L(M(f))-M(L(f))=0$ implies that 
the Yang equation \eqref{yang} holds. 
Here the existence of $N$-independent solutions of the linear system \eqref{lin_yang} is an assumption, however, we will show in section 3.2 that it actually exists for the soliton solution cases. Then the solution $f$ can be rewritten as an $N\times N$ matrix which consists of the $N$-independent solutions 
as column vectors of length $N$.

The Darboux transformation is an auto-B\"acklund transformation of the linear system (\ref{lin_yang}). 
Firstly, we assume that an initial solution $J$ of the Yang equation, and a solution $f=f(\zeta)$ of the linear system \eqref{lin_yang} are given. In order to define the Darboux transformation, we prepare a special solution $\theta(\Lambda):=f(\Lambda)$ for a fixed spectral parameter matrix $\zeta=\Lambda$. 
Then we can define the Darboux transformation as follows: 
\begin{eqnarray}
\label{Darboux_phi}
D_{\Lambda}: \left\{
\begin{array}{l}
f\mapsto 
f^\prime=
f \zeta - \theta \Lambda \theta^{-1} f, 
\\
J\mapsto 
J^\prime= -\theta \Lambda \theta^{-1} J
\end{array}
\right.
\label{Darboux_J}
\end{eqnarray}
which keeps the linear system \eqref{lin_yang} invariant in form, that is, 
\begin{eqnarray}
L^\prime(f^\prime)&:=&
J^\prime \partial_{w}(J^{\prime -1} f^\prime)
- (\partial_{\widetilde{z}}f^\prime)\zeta=0,\nonumber\\
M^\prime(f^\prime)&:=&
J^\prime \partial_{z}(J^{\prime -1} f^\prime)
- (\partial_{\widetilde{w}}f^\prime)\zeta=0.
\end{eqnarray}
As mentioned before, the transformation \eqref{Darboux_phi} becomes trivial if the spectral parameter is a scalar matrix because $\Lambda$ commutes with $\theta$ in this case.

By applying $n$ iterations of the Darboux transformation to a trivial seed solution $J=1$, we can get a nontrivial solution $J=D_{\Lambda_n}\circ \cdots \circ D_{\Lambda_1}(1)$ which is represented in terms of the quasideterminants \cite{GeRe} in a compact form \cite{GNO, GHHN}:
\begin{eqnarray}
\label{Jn}
J=
\left|
\begin{array}{cccc}
\theta_1&\cdots&\theta_n& 1\\
\theta_1\Lambda_1&\cdots &\theta_n\Lambda_n& 0\\
\vdots   && \vdots& \vdots\\
\theta_1\Lambda_1^{n-1}&\cdots&\theta_n\Lambda_n^{n-1}& 0\\
\theta_1\Lambda_1^{n}&\cdots& \theta_n\Lambda_n^{n}& \fbox{$0$}
\end{array}\right|, 
\end{eqnarray}
where $0$ and $1$ are $N\times N$ zero and identity matrices respectively, 
and each $\theta_j=\theta_j(\Lambda_j)~(j=1,2,\cdots,n)$ is a solution ($N \times N$ matrix) of the initial linear system with $J =1, \zeta=\Lambda_j$: 
\begin{eqnarray}
\label{chasing}
\partial_w \theta_j=(\partial_{\widetilde{z}}\theta_j)\Lambda_j,~~~
\partial_z \theta_j=(\partial_{\widetilde{w}}\theta_j)\Lambda_j.
\end{eqnarray}
For details of quasideterminants, please refer to e.g. \cite{GGRW,Huang}. 
In this paper, we give a definition of the quasideterminant 
only for the following type of matrix: 
\begin{eqnarray*}
\label{2x2}
\begin{vmatrix}
M\!\!\!\!\!&\begin{array}{cc}C_1&C_2\end{array}\\
\begin{array}{c}
R_1\!\!\!\!\!\\R_2\!\!\!\!\!
\end{array}
&\fbox{$
	\begin{array}{cc}
	a&b\\c&d
	\end{array}$}
\end{vmatrix}
:=
\left(\begin{array}{cc}a&b\\c&d\end{array}\right)
-
\left(
\begin{array}{c}
\!\!R_1\!\!\\\!\!R_2\!\!
\end{array}
\right)
M^{-1}
\left(
\begin{array}{cc}\!\!C_1\!&\!C_2\!\!\end{array}
\right)
=
\dfrac{1}{\vert M\vert}
\left(\begin{array}{cc}\begin{vmatrix}
M&C_1\\R_1&a
\end{vmatrix}&\begin{vmatrix}
M&C_2\\R_1&b
\end{vmatrix}\smallskip \\\begin{vmatrix}
M&C_1\\R_2&c\\
\end{vmatrix}&\begin{vmatrix}
M&C_2\\R_2&d\\
\end{vmatrix}\end{array}\right), 
\end{eqnarray*}
where $a,b,c,d$ are $1\times 1$ elements, and $C_1, C_2$ are $k$-component column vectors, and $R_1$ and $R_2$ are $k$-component row vectors, and $M$ is a $k\times k$ matrix for any $k\in \mathbb{N}$. 

For our purpose in this paper, we only discuss $N=2$ case in the following sections.

\subsection{Soliton Solutions for $G=U(2)$}

In this subsection, we present the soliton solutions given by \cite{HaHu2}. 
An example of the multi-soliton solution for $G=U(2)$ is given by solving the equation (\ref{chasing}): 
\begin{eqnarray}
\label{n-soliton solution}
\label{CS_n}
\theta_j=\left(
\begin{array}{cc}
e^{L_j}
& 
e^{-\overline{L}_j}
\\ 
-e^{-L_j}
& 
e^{\overline{L}_j}
\end{array}\right)
,~~~
\Lambda_j=
\left(
\begin{array}{cc}
\lambda_j & 0 \\
0 & \mu_j
\end{array}
\right),
\end{eqnarray}
where the two kinds of spectral parameters 
$\lambda_j, \mu_j$ ($j=1,2,\cdots,n$) are complex constants with the following mutual relationship on each real space:
\begin{eqnarray}
(\lambda_j, \mu_j)&\!\!\!=\!\!\!&
\left\{
\begin{array}{l}
(\lambda_j, \overline{\lambda}_j) ~~ \mbox{on}~ \mathbb{U}, 
\smallskip \smallskip \\
(\lambda_j, -1 / \overline{\lambda}_j) ~~ \mbox{on} ~ \mathbb{E}.
\end{array} \label{(lambda_j, mu_j)}
\right.
\end{eqnarray}
In this paper, we put a further condition $\vert \lambda_j \vert=1$ for all $j$ 
so that the group $G$ is unitary. 
The powers $L_j$ of the exponential function are linear in the complex coordinates: 
\begin{eqnarray}
L_j:=\lambda_j\alpha_jz+\beta_j\widetilde{z}+\lambda_j\beta_jw+\alpha_j\widetilde{w},~~~\alpha_j, \beta_j \in \mathbb{C}. 
\end{eqnarray}

\noindent
\textbf{Remark 1:}
The determinant of the $n$-soliton solution $J$ is constant \cite{HaHu2, Huang}:
\begin{eqnarray}
\label{Determinant of n-soliton solution}
\left|J\right|=\prod_{j=1}^{n}\lambda_j\mu_j.
\end{eqnarray}

\medskip
\noindent
\textbf{Remark 2:}
$\theta_j$ can be decomposed into, for instance
\begin{eqnarray}
\label{XTheta}
\theta_j
=\left(
\begin{array}{cc}
e^{L_j} & e^{-\overline{L}_j}
\\ 
-e^{-L_j} & e^{\overline{L}_j}
\end{array}\right)
=\left(
\begin{array}{cc}
e^{X_j}  &   e^{i\Theta_j}\\ 
-e^{-i\Theta_j}  &  e^{X_j} 
\end{array}\right)
\left(
\begin{array}{cc}
e^{-\overline{L}_j} & 0  \\ 
0  &  e^{-L_j} 
\end{array}\right), 
\end{eqnarray}
where $X_j:=L_j+\overline{L}_j$, 
$i\Theta_j:=L_j-\overline{L}_j$. 
The second factor diag$(e^{-\overline{L}_j},e^{-L_j})$
can be eliminated in the $n$-soliton solutions (\ref{Jn})
due to a property of the quasideterminant.
Hence the $n$-soliton solutions (\ref{Jn}) 
depend only on $X_j$ and $\Theta_j$.

\section{Action Density for Soliton Solutions}

In this section, we give a brief summary of soliton solutions in $G=U(2)$ 4dWZW model \cite{HHK}. The explicit results of action density for one- and two-soliton are summarized in section 4.1. The asymptotic analysis of action density for $n$-soliton solutions are summarized in 4.2.

\subsection{One-Soliton and Two-Soliton Solutions}

The action densities for the one-soliton solution (\eqref{n-soliton solution} for $j=1$) are as follows:
\begin{eqnarray}
\label{L_sigma}
{\cal{L}}_\sigma(x)
&=&
\frac{1}{8\pi}
d_{11}~\!\mbox{sech}^2 X_1,\\
{\cal{L}}_{\scriptsize{\mbox{WZ}}}(x)&=&0,
\label{L_WZ}
\end{eqnarray}
where the constant $d_{11}$ is given in Table \ref{Table_1} and clearly a real number which implies ${\cal{L}}_\sigma$ is real-valued . 
The peak of the action density lies on the three-dimensional hyperplane described by the linear equation $X_1=0$ on each real space. Hence  we call soliton of this type (codimension-one solitons) the soliton wall which  can be considered as a higher-dimensional analogue of lower-dimensional solitons, such as the KP solitons. 
We note that the action density vanishes identically in the case of $\alpha_1, \beta_1, \lambda_1 \in \mathbb{R}$ on $\mathbb{U}$. 
All the above results also hold in the case of $\mathbb{E}$. 

\begin{table}[h]
\!\!\!\!\caption{Summary of Coefficients}
\label{Table_1}
\medskip
\begin{center}
\begin{tabular}{|c|c|c|}
		\hline
		Space & $\mathbb{U}$ & $\mathbb{E}$   \\ 
		(signature) & $(+,+,-,-)$ & $(+,+,+,+)$   \\
		\hline
		\hline
		$a \in \mathbb{R^{+}}$ & $\left| \lambda_1 - \lambda_2 \right|^2 \textgreater 0$ & 
$\left| \lambda_1 - \lambda_2 \right|^2\textgreater 0$ \\ 
		\hline 
		$b \in \mathbb{R}$ & $\left| \lambda_1 - \overline{\lambda}_2 \right|^2 \textgreater 0$ 
& $-\left| \lambda_1\overline{\lambda}_2 +1 \right|^2 \textless 0$ \\
		\hline 
		$c \in \mathbb{R}$ & $\left( \lambda_1 - \overline{\lambda}_1 \right)\!\!\left( \lambda_2 - \overline{\lambda}_2 \right)$ 
& $\left( \left| \lambda_1 \right|^2 + 1 \right)\!\!\left( \left| \lambda_2 \right|^2 + 1 \right)$ \\ 
		\hline
		$d_{jk}$  & $\underline{\left( \alpha_j\overline{\beta}_k - \beta_j\overline{\alpha}_k \right)
			\!\left( \lambda_j - \overline{\lambda}_k \right)^3}$ 
&  $\underline{\left( \alpha_j\overline{\alpha}_k + \beta_j\overline{\beta}_k \right)
			\!\left( \lambda_j\overline{\lambda}_k + 1 \right)^3}$  \\ 
		$(=\overline{d}_{kj})$ & $\lambda_j\overline{\lambda}_k$ 
& $\lambda_j\overline{\lambda}_k$ \\ 
		\hline
		$e_{jk}$ &  $\underline{(\alpha_j\beta_k - \beta_j\alpha_k)( \lambda_j - \lambda_k)^3}$ 
& $\underline{(\alpha_j\beta_k - \beta_j\alpha_k)( \lambda_j - \lambda_k)^3}$ \\
		& $\lambda_j\lambda_k$ 
& $\lambda_j\lambda_k$  \\
\hline 
\end{tabular}
\end{center}
\end{table}

The result of two-soliton (\eqref{n-soliton solution}, $j$=1, 2.) becomes more non-trivial and the corresponding NL$\sigma$M action density is obtained as follows:
\begin{eqnarray}
{\cal{L}}_\sigma(x)=
\label{NL Sigma term_2-Soliton_form 1}
\frac{
	\left\{ 
	\begin{array}{l}
	~ab
	\left[
	d_{11}\cosh^2 X_2 + d_{22}\cosh^2 X_1
	\right]
	\medskip \\
	\!\!+ ac\left[
	d_{12}~\!\displaystyle{\cosh^2 \left(\frac{X_1 + X_2 - i\Theta_{12}}{2} \right)}
	+ d_{21}~\!\displaystyle{\cosh^2 \left(\frac{X_1 + X_2 + i\Theta_{12}}{2} \right)}
	\right]
	\medskip \\
	\!\!- bc\left[
	e_{12}~\!
	\displaystyle{\sinh^2 \left(\frac{X_1 - X_2 - i\Theta_{12}}{2}
		\right)}
	+\overline{e}_{12}~\!
	\displaystyle{\sinh^2 \left(\frac{X_1 - X_2 + i\Theta_{12}}{2} \right)}
	\right]
	\end{array}
	\!\!\!\right\}
}
{2\pi\displaystyle{
		\left[
		a\cosh(X_1 + X_2)
		+
		b\cosh(X_1 - X_2)
		+
		c\cos \Theta_{12}
		\right]^2}
	}~~~
\end{eqnarray} 
where $a, b, c, d_{jk}, e_{jk}$ are defined in the Table \ref{Table_1} for each real space. (The difference between $\mathbb{E}$ and $\mathbb{U}$ appears only in the coefficients like the one-soliton case.) 
Note that the coefficients in Table \ref{Table_1} also guarantee the NL$\sigma$M action density to be real-valued on $\mathbb{U}$ 
and $\mathbb{E}$. We can prove that the action density is everywhere nonsingular. 

The configuration \eqref{NL Sigma term_2-Soliton_form 1} can be interpreted as two intersecting one-soliton walls in the asymptotic region. The asymptotic analysis of the NL$\sigma$M action density \eqref{NL Sigma term_2-Soliton_form 1} gives the following results (where double-sign corresponds):
\begin{eqnarray}
\label{Asymptotic_2-Soliton}
-8\pi {\cal{L}}_\sigma (x)
 \longrightarrow 
\left\{
\begin{array}{l}
(1)~X_1 ~\mbox{is finite}, ~X_2  \rightarrow  \pm \infty :~
d_{11}~\!{\mathrm{sech}^2\left( X_1 \pm \delta \right)}
\medskip \\
(2)~X_2 ~\mbox{is finite}, ~X_1  \rightarrow  \pm\infty :~  
d_{22}~\!{\mathrm{sech}^2\left( X_2 \pm \delta \right)}
\end{array},
\right.
\end{eqnarray}
where the position shift factor is 
\begin{eqnarray}
\delta
:= \frac{1}{2}\log \frac{(\lambda_1-\lambda_2)(\mu_1-\mu_2)}{(\lambda_1-\mu_2)(\mu_1-\lambda_2)}.
\end{eqnarray}
This shift, called the phase shift, results from the non-linear soliton interaction which is a common phenomenon of solitons in lower dimensions.
The Wess-Zumino action density for the two-soliton is much more complicated, however, it is non-singular and decays to zero exponentially in any direction of the asymptotic region. We make a conjecture that the Wess-Zumino action $S_{\scriptsize{\mbox{WZ}}}=0$.

\subsection{ $n$-Soliton Solutions (Asymptotic Analysis)}
 
Finally, we show the asymptotic behavior of $n$-soliton in the sense of action density on $\mathbb{U}$. Without loss of generality, we consider an asymptotic region ${\mathscr{R}}_K$ in which $X_K$ is finite and $r:=(x^1)^2+(x^2)^2+(x^3)^2+(x^4)^2\rightarrow\infty$ for given $K \in \left\{ 1, 2, \cdots, n \right\}$. The NL$\sigma$M action density of $n$-soliton is now asymptotic to one-soliton
:
\begin{eqnarray}
\label{action density_asym}
{\cal{L}}_\sigma(x) =
-\frac{1}{16\pi}\mathrm{Tr}
\left[
(\partial_{\mu}J)J^{-1}  
(\partial^{\mu}J)J^{-1} \right]
\stackrel{{\mathscr{R}}_K}{\simeq}
-\frac{1}{8\pi} d_{KK}~\!\mbox{sech}^2\left( X_K + \delta_K \right),
\end{eqnarray}
where $d_{KK}$ is defined in Table \ref{Table_1} and the phase shift factor is
\begin{eqnarray}
\label{phase shifts}
\delta_K 
\stackrel{\mathbb{U}}{=}
\displaystyle{\sum_{j=1, j \neq K}^{n}\log\left| \frac{\lambda_K - \lambda_j}{\lambda_K - \overline{\lambda}_j} \right|}.
\end{eqnarray}
Since the result of \eqref{action density_asym} is valid for arbitrary $K$ in $\left\{ 1, 2, \dots, n\right\}$,
we can regard the behavior of (not proved in fact) the  $n$-soliton as a ``non-linear superposition'' of $n$  mutually nonparallel one-solitons 
and each one-soliton (dominated by $X_K$) maintains its form invariant but is shifted by $\delta_K$ in the asymptotic region $\mathscr{R}_K$.

In conclusion, in the asymptotic region the $n$-soliton solution possesses $n$ isolated and localized lumps of the NL$\sigma$M action density, and we can interpret it as $n$ intersecting soliton walls. The phase shift factors are also obtained explicitly. The scattering process of the $n$-soliton solution is quite similar to that of the KP solitons \cite{OhWa, Kodama}. 
On the other hand, the Wess-Zumino action density identically vanishes in the asymptotic region. 

\medskip
\noindent
{\bf Remark 3:} We can define a WZW model in $(1+2)$ dimensional by dimensional reduction and construct codimension-one soliton solutions (soliton walls) of it. In this reduced model, we can define the Hamiltonian (energy) density via the Legendre transformation. We proved that the energy density for the Wess-Zumino term identically vanishes and therefore the total energy density depends only on the NL$\sigma$M term and is positive definite for $G=U(N)$. This is the reason why we always focus our discussion on the NL$\sigma$M term in this paper. Furthermore, the total energy density is localized on a two-dimensional plane in $(1+2)$-dimensions whose peak coincides with that of the action density and quite similar to the KP solitons. 
For the detailed discussion, please refer to section 6 in \cite{HHK}.

\section{Action Density for Instanton Solutions}

The main purpose of this section is to discuss instanton solutions of the 4dWZW model on the Euclidean space $\mathbb{E}$. 
In section 5.1, we review some preknowledge of the ASDYM equation and the Yang equation, and then discuss the equivalence between the two equations. In section 5.2, we construct instanton solutions in the 4dWZW model for $G=SL(2,\mathbb{C})$ which are new results.

\subsection{ASDYM Equation and Yang Equation}

Before our main discussion, we introduce some preknowledge of anti-self-dual Yang-Mills equation on the Euclidean space $\mathbb{E}$. 
Here we use $G_{\scriptsize{\mbox{YM}}}$ to denote Yang-Mills gauge group which is in general different from the group $G$ in the 4dWZW model. For our purpose in this paper, we only consider the case $G_{\scriptsize{\mbox{YM}}}=GL(N,\mathbb{C})$ or subgroups of it. Then gauge fields $A_\mu$ takes values in the Lie-algebra of $G_{\scriptsize{\mbox{YM}}}$. Especially for $G_{\scriptsize{\mbox{YM}}}=U(N)$, the gauge fields $A_\mu$ are anti-Hermitian matrix. 

Covariant derivatives and field strengths are defined respectively by $D_\mu:=\partial_\mu+A_\mu$ and by $F_{\mu\nu}:=[D_\mu,D_\nu]=\partial_\mu A_\nu-\partial_\nu A_\mu+[A_\mu,A_\nu]$. The ASDYM equation is defined by the anti-self-duality in the sense of the Hodge dual operator $*$: $F_{\mu\nu}=-*F_{\mu\nu}$ which can be rewritten in the complex representation:
\begin{eqnarray}
F_{wz}=0,~~~
F_{\overline{w}\,\overline{z}}=0,~~~
F_{z\overline{z}}+F_{w\overline{w}}=0.
\label{asdym}
\end{eqnarray}
A Lax representation of the ASDYM equation is given by 
\begin{eqnarray}
L(f)&:=&D_w f -  D_{\overline{z}}f\zeta =
0,\nonumber\\
M(f)&:=&D_z f +  D_{\overline{w}}f \zeta =
0.
\label{lin_asdym}
\end{eqnarray}
Here $\zeta$ is an $N\times N$ constant matrix, called the spectral parameter matrix. 
The compatibility condition $L(M(f))-M(L(f))=0$ of the linear system \eqref{lin_asdym} gives rise to the ASDYM equation \eqref{asdym}. 

Let $g$ be a $G$-valued function. The local transformation $D_\mu\mapsto g^{-1} D_\mu g, f\mapsto g^{-1} f$ is called the gauge transformation. The gauge transformation acts on the linear system \eqref{lin_asdym} covariantly and on the gauge fields and the field strengths as the following map:
\begin{eqnarray}
\label{gauge}
A_\mu\mapsto g^{-1} A_\mu g+g^{-1} \partial_\mu g,~~~
F_{\mu\nu}\mapsto g^{-1} F_{\mu\nu} g.
\end{eqnarray}
Moduli space of the ASD gauge fields ${\cal M}_{\scriptsize{\mbox{ASDYM}}}$ can be defined as the solution space of the ASDYM equation \eqref{asdym} up to the gauge transformation \eqref{gauge}. 

In order to discuss equivalence between the ASDYM equation and the Yang equation, let us define a map from a solution of the former to a solution of the latter.
Here we assume that the spectral parameter matrix $\zeta$ is a diagonal matrix. We note that the first equation of the ASDYM equation
(\ref{asdym}) is the compatible condition of the linear system $D_z h=0, D_w h=0$. This can be interpreted as the limit of \eqref{lin_asdym} that all diagonal elements go to zero. 
Here we assume that there exists $N$-independent solutions of this linear system so that $h$ can be considered as an $N\times N$ regular matrix whose columns consist of the $N$ independent solutions. (If $\zeta$ is a scalar matrix, the existence is guaranteed by the Frobenius theorem.) 
Similarly, from the second equation of the ASDYM equation (\ref{asdym}), there exists an $N\times N$ regular matrix $\widetilde{h}$ which is a solution of the linear system $D_{\overline{z}} \widetilde{h}=0, D_{\overline{w}} \widetilde{h}=0$. (This can be interpreted as the limit of \eqref{lin_asdym} that all diagonal elements go to infinity.) 
By defining a new matrix $J:=\widetilde{h}^{-1}h$,
the third equation of the ASDYM equation (\ref{asdym}) becomes the Yang equation \eqref{Yang} in $\mathbb{E}$, which is proved without gauge fixing by showing: $\partial_{\overline{z}}((\partial_z J) J^{-1})=\widetilde{h}^{-1}F_{z\overline{z}}\widetilde{h}$ \cite{Huang}.

We can define a map from the a solution $J$ of the Yang equation \eqref{yang} to a solution $A_m$ of the ASDYM equation \eqref{asdym} by decomposing $J$ into two $N\times N$ regular matrices $h$ and $\widetilde{h}$ so that $J=\widetilde{h}^{-1}h$. Then we obtain the ASD gauge fields in terms of $h$ and $\widetilde{h}$ as follows (Note that $D_z h=0 \Leftrightarrow A_{z}=-(\partial_z h) h^{-1}$ and so on.): 
\begin{eqnarray}
\label{A}
A_{z}=-(\partial_z h) h^{-1}, ~A_w=-(\partial_w h) h^{-1},~
A_{\overline{z}}=-(\partial_{\overline{z}}\widetilde{h}) \widetilde{h}^{-1}, ~
A_{\overline{w}}=(\partial_{\overline{w}}\widetilde{h}) \widetilde{h}^{-1}. 
\end{eqnarray}
By taking the gauge fixing: $\widetilde{h}=1\Rightarrow A_{\overline{z}}=A_{\overline{w}}=0$, the linear system \eqref{lin_asdym} coincides with \eqref{lin_yang}. (Under this gauge, gauge fields are represented in terms of $J$ such as $A_z=-(\partial_z J) J^{-1}$ because of $h=J$.) 
We can prove that unitary gauge group $G_{\scriptsize{\mbox{YM}}}=U(N)$ is realized by the condition that $h^\dagger \widetilde{h}= \widetilde{h} h^\dagger=1$ \cite{Yang, Huang}. In this case, $J=h^\dagger h$ is hermitian. 

Another moduli space ${\cal M}_{\scriptsize{\mbox{Yang}}}$ can be defined as the solution space of the Yang equation \eqref{yang} up to the holomorphic/anti-holomorphic transformation \eqref{PQ}. The gauge transformation \eqref{gauge} acts on $h$ and $\widetilde{h}$ as $h\mapsto g^{-1} h,~ \widetilde{h}\mapsto g^{-1} \widetilde{h}$, and hence leaves $J$ as it is. 
On the other hand, the holomorphic/anti-holomorphic transformation \eqref{PQ} acts on $h$ and $\widetilde{h}$ as $h\mapsto hQ(\overline{z},\overline{w}),~ \widetilde{h}\mapsto \widetilde{h}P(z,w)^{-1}$, and hence leaves $A_m$ as it is. (See \eqref{A}.) 
We note that there is ambiguity in the choice of the solutions $h$ and $\widetilde{h}$ of the linear systems in the map from $A_m$ to $J$, that is, if $h$ is a solution of the linear system $D_zh=0, D_w h=0$, another $h^\prime:=hQ~(Q(\overline{z},\overline{w})\in GL(N,\mathbb{C}))$ also satisfies the linear system, however, this ambiguity can be absorbed by the degree of freedom of the anti-holomorphic transformation. There is also ambiguity in the choice of the decomposition of $J$ into $h$ and $\widetilde{h}$ in the map from $J$ to $A_m$, that is, if $h, \widetilde{h}$ is a solution in the decomposition $J=\widetilde{h}^{-1}h$, another $h^\prime:=gh, \widetilde{h}^\prime:=g\widetilde{h}~(g(x)\in GL(N,\mathbb{C}))$ also satisfies $J=\widetilde{h}^{\prime -1}h^\prime$, however, this ambiguity can be absorbed by the degree of freedom of the gauge transformation of $g$. 
Therefore we can conclude that there is one-to-one correspondence between ${\cal M}_{\scriptsize{\mbox{ASDYM}}}$ and ${\cal M}_{\scriptsize{\mbox{Yang}}}$. This one-to-one correspondence holds in $\mathbb{U}$ as well. 

\subsection{Instanton Solutions for Instanton Solutions for 
$G=SL(2,\mathbb{C})$}

Let us construct $G=SL(2,\mathbb{C})$ instanton solutions which correspond to
$G_{\scriptsize{\mbox{YM}}}=SU(2)$ instantons in the Yang-Mills side. 
The matrix $J$ should be hermitian because of the reason below Eq. \eqref{A} and $\det J=1$ by assumption. Then it can be uniquely parametrized as follows:
\begin{eqnarray}
\label{Yang's J-matrix}
J
=\frac{1}{\varphi}
\left(
\begin{array}{cc}
1 & \overline{\rho} \\
\rho  & \varphi^2 + \rho \overline{\rho}
\end{array}
\right), 
\end{eqnarray}
where $\varphi$ is a real-valued function and $\rho$ is a complex-valued function. Then the Yang equation can be represented in terms of $\varphi$ and $\rho$. 
In order to solve this, let us put the following ansatz \cite{Yang}:
\begin{eqnarray}
\label{relation of phi and rho}
\partial_{z}\varphi = - \partial_{\overline{w}}\overline{\rho}, ~ 
\partial_{w}\varphi =   \partial_{\overline{z}}\overline{\rho}, ~
\partial_{\overline{z}}\varphi = - \partial_{w}{\rho}, ~ 
\partial_{\overline{w}}\varphi =   \partial_{z}{\rho}. 
\end{eqnarray}
Then the Yang equation reduces to the four-dimensional Laplace equation for $\varphi$: 
\begin{eqnarray}
\label{Laplace eq}
(\partial_{z} \partial_{\overline{z}} + \partial_{w}\partial_{\overline{w}})\varphi =0,
\end{eqnarray}
A fundamental solution of \eqref{Laplace eq} is:
\begin{eqnarray}
\label{tHft 1-instanton solution}
\varphi=
1+\frac{\lambda^2}{z\overline{z} + w\overline{w}}, 
\end{eqnarray}
where $\lambda$ is a real constant. 
This actually leads to a class of 't Hooft one-instanton solutions.

\medskip
\noindent
{\bf Remark 4:} The $SU(2)$ ASD gauge fields are obtained from the solution by decomposing $J$ into $h$ and $\widetilde{h}$ so that $J=\widetilde{h}^{-1}h$, $\widetilde{h}^\dagger=h^{-1}$ and $\det J=1$. If we assume further that $h$ is lower-triangular and $\widetilde{h}$ is upper-triangular, this decomposition is uniquely determined and a gauge is fixed. This is called Yang's R-gauge. Under this gauge and the ansatz \eqref{relation of phi and rho}, the gauge fields can be represented in terms of $\varphi$ only and rewritten in the form of the 't Hooft ansatz. (A detailed derivation can be referred to e.g. in \cite{Huang}.) The Yang-Mills action density for \eqref{tHft 1-instanton solution} is calculated as follows:
\begin{eqnarray}
\label{YMaction}
{\cal{L}}_{\scriptsize{\mbox{YM}}}(x)
:=-\frac12\mbox{Tr}F_{mn}F^{mn}
=2\mbox{Tr}\left( F_{z\overline{z}}^2 + F_{z\overline{w}}F_{w\overline{z}}\right)
=\frac{12\lambda^4}{(\lambda^2 + z\overline{z} + w\overline{w})^4},
\end{eqnarray}
We note that the action density is nonsingular everywhere while the gauge fields and field strengths are not. The singularities in the gauge fields and field strengths can be eliminated by a singular gauge transformation. 

\medskip
In order to calculate the 4dWZW action density for \eqref{tHft 1-instanton solution}, we have to find  a solution $\rho$ such that the differential equations  \eqref{relation of phi and rho}  are satisfied.  In fact, such solution $\rho$ is not unique. For instance, the following $\rho_1$ and $\rho_2$ are solutions of \eqref{relation of phi and rho}:
\begin{eqnarray}
\rho_1:=\frac{\lambda^2 w}{(z\overline{z} + w\overline{w})\overline{z}},~~~
\rho_2:=\frac{-\lambda^2 z}{(z\overline{z} + w\overline{w})\overline{w}}.
\end{eqnarray}
Here we can take a linear combination of $\rho_1, \rho_2$ as a new solution of \eqref{relation of phi and rho}: 
\begin{eqnarray}
\label{rho_tHft 1-instanton}
\rho
= \frac12 (\rho_1+\rho_2)=
\frac{\lambda^2(w\overline{w}-z\overline{z})}{2(z\overline{z} + w\overline{w})\overline{z}\overline{w}}. 
\end{eqnarray}	
By substituting \eqref{tHft 1-instanton solution} and \eqref{rho_tHft 1-instanton} into the following formula:  (Here the assumption of constant determinant for $J$ is necessary.)
\begin{eqnarray}
\label{TrA^2}
\mbox{Tr}(Y_mY_n)
&\!\!\!\!=\!\!\!\!&\frac{-1}{\varphi}
\left[
2(\partial_{m}\varphi)(\partial_{n}\varphi) + (\partial_{m}\rho)(\partial_{n}\overline{\rho})+ (\partial_{m}\overline{\rho})(\partial_{n}\rho)
\right],
\\
\label{TrA^3}
\mbox{Tr}(Y_m Y_n Y_p)
&\!\!\!\!=\!\!\!\!&
\frac{1}{\varphi^3}
\left|
\begin{array}{ccc}
\partial_{m}\overline{\rho} & \partial_{m}\rho & \partial_{m}\varphi \\
\partial_{n}\overline{\rho} & \partial_{n}\rho & \partial_{n}\varphi \\
\partial_{p}\overline{\rho} & \partial_{p}\rho & \partial_{p}\varphi
\end{array}
\right|, 
\end{eqnarray}
the NL$\sigma$M action density and
the Wess-Zumino action density are
 calculated as follows:
\begin{eqnarray}
\nonumber
{\cal{L}}_{\sigma}(x)&=&
{\mbox{Tr}}({Y_zY_{\overline{z}} + Y_{\overline{z}}Y_z +Y_wY_{\overline{w}} + Y_{\overline{w}}Y_w})
=
\frac{-\lambda^4(z\overline{z} + w\overline{w})^3}{2(z\overline{z}w\overline{w})^2(\lambda^2 + z\overline{z} + w\overline{w})^2},\\
{\cal{L}}_{\scriptsize{\mbox{WZ}}}(x)&=&{\mbox{Tr}}(Y_zY_wY_{\overline{w}})z +
{\mbox{Tr}}(Y_{\overline{z}}Y_wY_{\overline{w}})\overline{z} +
{\mbox{Tr}}(Y_wY_zY_{\overline{z}})w +
{\mbox{Tr}}(Y_{\overline{w}}Y_zY_{\overline{z}})\overline{w}
 \nonumber\\
&\!\!\!\!=\!\!\!\!&
\frac{-\lambda^8(z\overline{z} + w\overline{w})(z\overline{z} - w\overline{w})^2}{2(z\overline{z}w\overline{w})^2(\lambda^2 + z\overline{z} + w\overline{w})^4},
\label{WZWaction}
\end{eqnarray}
where $Y_m:=(\partial_m J)J^{-1}$. 
We can find that both of the action densities have a peak at the origin and 
are asymptotic to zero at infinity. This implies that this solution represents an instanton in the 4dWZW model. However there exists singular locus specified by $z\overline{z}w\overline{w}=0$ which results from the singularities in $\rho$  (Cf: \eqref{rho_tHft 1-instanton}).  These singularities cannot be removed by any gauge transformation \eqref{gauge} because $J$ is gauge invariant. It is a future work to study whether a suitable (anti)holomorphic transformation \eqref{PQ} can remove it. 



\section{Ward Conjecture}

In this section, we give some known examples of the Ward conjecture. The first example is the KdV equation which is derived from the ASDYM equation \cite{MaSp,MaWo}. The second and third examples are the sine-Gordon and Liouville equations which are derived from the Yang equation \cite{GWW}. 

\subsection{Reduction to KdV equation from ASDYM equation}

Let us start with the $SL(2,\mathbb{C})$ ASDYM equation \eqref{asdym} and take a dimensional reduction by putting translational invariance along 
$X=\partial_w-\partial_{\widetilde{w}}, Y=\partial_{\widetilde{z}}$. 
Then the fields depend on two variables $(t,x)\equiv (z,w+\widetilde{w})$ and  
$\varPhi_X:=A_w-A_{\widetilde{w}}$, 
$\varPhi_{\widetilde{z}}:=A_{\widetilde{z}}$
can be considered as Higgs fields. The ASDYM eq. \eqref{asdym} becomes 
\begin{eqnarray}
\label{asdym2}
&&\partial_x\varPhi_{\widetilde{z}}
 +[A_{\widetilde{w}},\varPhi_{\widetilde{z}}] =0,~~
 \partial_t\varPhi_{\widetilde{z}}
 +\partial_x A_{w} -\partial_x A_{\widetilde{w}}
 +[A_z,\varPhi_{\widetilde{z}}]
 -[A_{w},A_{\widetilde{w}}]=0,~~\\
&&
 \partial_x A_z -\partial_t A_w+[A_w,A_z] =0.
\label{asdym3}  
 \end{eqnarray}
In order to get the KdV equation, we put the following reduction condition 
for the gauge fields: 
\begin{eqnarray}
 \label{kdv}
A_{\tilde{w}}\!=\!\left(\begin{array}{cc}\!0&0\!\\\!u/2&0\!\end{array}\right), 
A_{\tilde{z}}\!=\!\left(\begin{array}{cc}\!0&0\!\\\!1&0\!\end{array}\right),
A_w\!=\!\left(\begin{array}{cc}\!0&-1\!\\\!u&0\!\end{array}\right),
A_z\!=\!\frac{1}{4}
\left(\begin{array}{cc}\! \partial_x u& -2u \!\\
\! \partial_x^2 u+2u^2 &-\partial_x u\!
\end{array}\right).
\end{eqnarray}
Under this condition, the reduced ASDYM eq. (\ref{asdym2}), (\ref{asdym3})  automatically satisfied except for the $(2,1)$ component of (\ref{asdym3}). The equality holds if we take the following constraint:
\begin{eqnarray}
\label{red}
  \partial_t u=\frac{1}{4} \partial_x^3u
  +\frac{3}{2}u \partial_x u,
\end{eqnarray}
which is the KdV equation.
We note that the variable $t$ in the KdV equation \eqref{red} is the time coordinate and hence it must be real. In the split signature, this coordinate $t=z$ can be real as in \eqref{U} while in the Euclidean signature, this $t=z$ must be complex as in \eqref{E}.  Hence the reduced equation \eqref{red} cannot be considered as the standard KdV equation living in $(1+1)$ dimensional (real) space-time  if we consider the Euclidean signature. This is the reason why we usually consider the split signature in the context of the Ward conjecture. 

\medskip
\noindent
{\bf Remark 5:}  In fact, we can show that the above reduction conditions   
(\ref{kdv}) result from the standard Lax representation of the KdV equation:
\begin{eqnarray}
P \psi=
(\partial_x^2+u) \psi=
\lambda \psi,
 ~~~B \psi=
\left(\partial_t-\partial_x^3-\frac{3}{2}u\partial_x-\frac{3}{4}(\partial_x u)\right) \psi
=0,
 \label{Lax}
\end{eqnarray}
where the compatibility condition $[P,B]=0$ gives rise
to the KdV equation \eqref{red}. 
In order to rewrite the scalar linear system \eqref{Lax} into the $2\times 2$ form, we introduce a two component wave function: 
\begin{eqnarray}
 \Psi=\left(\begin{array}{c}\psi\\\partial_x \psi\end{array}\right),
\end{eqnarray}
then the derivatives of $\Psi$ with respect to $x$ and $t$
are calculated from Eq. (\ref{Lax}) as follows 
\begin{eqnarray}
 \label{s}
 \partial_x \Psi&=&\left\{
 \left(\begin{array}{cc}0&1\\-u&0\end{array}\right)
 +\lambda\left(\begin{array}{cc}0&0\\1&0\end{array}\right)
 \right\}
\left(\begin{array}{c}\psi\\\partial_x \psi\end{array}\right),\\
  \partial_t\Psi&=&
   \left\{\frac{1}{4}
 \left(\begin{array}{cc}-\partial_x u&2u\\-\partial_x^2 u-2u^2&\partial_x u\end{array}\right)
 +\lambda\left(\begin{array}{cc}0&1\\-u/2&0\end{array}\right)
 +\lambda^2\left(\begin{array}{cc}0&0\\1&0\end{array}\right)
 \right\}
\left(\begin{array}{c}\psi\\\partial_x \psi\end{array}\right),\nonumber\\
&=&
   \left\{\frac{1}{4}
 \left(\begin{array}{cc}-\partial_x u&2u\\-\partial_x^2 u-2u^2&\partial_x u\end{array}\right)
 +\lambda\left[\partial_x+
                \left(\begin{array}{cc}0&0\\u/2&0\end{array}\right)\right]
 \right\}
\left(\begin{array}{c}\psi\\\partial_x \psi\end{array}\right).
\label{t}
\end{eqnarray}
It is nontrivial that
Eq. (\ref{t}) becomes linear in $\lambda$
as in the second line of Eq. (\ref{t}).
{}From this representation,
we can obtain the reduction condition (\ref{kdv})
by identifying the linear systems (\ref{lin_asdym}) with
Eqs. (\ref{s}) and (\ref{t}) 
together with $\partial_w=\partial_{\widetilde{w}}=\partial_x$ and $\partial_{\widetilde{z}}=0$.

\subsection{Reduction to sine-Gordon equation from Yang equation}

Let us discuss the reduction of the Yang equation (\ref{yang}) for $G=SL(2,\mathbb{C})$ to sine-Gordon equation \cite{GWW} 
by putting the following reduction condition for $J$:
\begin{eqnarray}
 J=e^{i\sigma_1 \widetilde{w}} g(z,\widetilde{z}) e^{i\sigma_1 w}.
\end{eqnarray}
Then the Yang equation (\ref{yang}) reduces to
\begin{eqnarray}
\label{yang2}
 \partial_{\widetilde{z}}\left( (\partial_{z} g) g^{-1}\right)
- \left[\sigma_1, ~g\sigma_1 g^{-1}\right]=0.
\end{eqnarray}
where Pauli matrices are defined as usual:
\begin{eqnarray}
 \sigma_1=
  \left(\begin{array}{cc}0&1\\1&0\end{array}\right),~
 \sigma_2=
  \left(\begin{array}{cc}0&-i\\i&0\end{array}\right),~
 \sigma_3=
  \left(\begin{array}{cc}1&0\\0&-1\end{array}\right).
\end{eqnarray}
By choosing $g=\exp\left[(i/2)\sigma_3 u\right]$, 
we get the sine-Gordon equation in the case of $\mathbb{U}$:
\begin{eqnarray}
 \partial_z\partial_{\widetilde{z}}u=4\sin u.
\end{eqnarray}

\subsection{Reduction to Liouville equation from Yang equation}

In the same way, we can obtain the Liouville equation from the Yang equation 
for $G=SL(2,\mathbb{C})$ \cite{GWW}. 
Let us put the reduction condition:
\begin{eqnarray}
 J=e^{\sigma_- \widetilde{w}}  g(z,\widetilde{z}) e^{-\sigma_+ w}.
\end{eqnarray}
Then the Yang equation (\ref{yang}) reduces to
\begin{eqnarray}
\label{yang2}
 \partial_{\widetilde{z}}
\left( (\partial_{z} g)g^{-1}\right)
- \left[\sigma_-, ~g\sigma_+ g^{-1}\right]=0,
\end{eqnarray}
where $\sigma_\pm:=(1/2)(\sigma_1\pm i\sigma_2)$. 
By choosing $g=\exp\left(\sigma_3 u\right)$, 
we get the Liouville equation in the case of $\mathbb{U}$:
\begin{eqnarray}
 \partial_z\partial_{\widetilde{z}}u=e^{2u},
\end{eqnarray}
We note that the Liouville equation can be derived from the $SU(2)$ ASDYM equation in $\mathbb{E}$ \cite{Witten}. Hence the reduction process is not unique. 

\section{Conclusion and Discussion}
 
In this paper, we  introduced soliton solutions of the $G=U(2)$ 4d WZW model by analyzing action densities for them. The behaviors of the soliton solutions are quite similar to the KP solitons,  more explicitly, one-soliton solutions are codimension-one solitons whose action densities are localized on three-dimensional hyperplanes in four dimensions. $n$-soliton solutions can be interpreted as a ``non-linear superposition'' of $n$ one-soliton solutions with phase shifts. In the split signature, our results imply that there exist codimension-one classical objects in the open $N=2$ string theory, and therefore, it is worth studying charges, mass/tension of the soliton solutions, and fluctuations around them in the context of the $N=2$ string theory. Furthermore, the N=2 string theory relates to mirror symmetries \cite{NeVa} and the M-theory \cite{KMO} and our solutions might play any roles in the corresponding situations. 

Classification of the soliton solutions would be important to clarify the moduli space so that we can perform path-integration in the background of the solitons. If successful, it might reveal non-perturbative effects in the open N=2 string theory.  A key point of the classification is the existence of Y-shape resonance solitons which are the building blocks of more complicated soliton interactions. Fortunately, the KP solitons are {already classified in terms of positive Grassmannians elegantly \cite{Kodama, KoWi} in the real-valued settings.  We anticipate that there is a quite similar description for solitons of ASDYM type because the dynamics of them are quite similar to the KP-solitons. Therefore, constructing Y-shape resonance solitons of  ASDYM type would be a good starting point toward the classification of ASDYM solitons.
On the other hand, codimension-one solitons of ASDYM type are also discussed in other contexts, e.g. \cite{deVega, Lee, LiZh, LQZ, LQYZ, Mason, Parkes}. The relationship between them and our solutions should be also clarified. 
It is also worth investigating other solutions such as rogue wave solutions (codimension-two solitons) \cite{Ohta}, rational solutions and elliptic solutions. 
These classical solutions are likely to be some objects in the open N=2 string theory as well. In order to confirm it, we need to calculate the corresponding action densities explicitly. 

Extension of our results to noncommutative (NC) spaces is also a significant topic for physics. It is a well-known fact that gauge theories in noncommutative spaces are equivalent to gauge theories (in commutative spaces) in the background of magnetic fields/B-fields \cite{CDS,DoHu, SeWi}. 
Completed instanton moduli spaces includes small instanton singularities \cite{Nakajima} which correspond to the size zero instantons, called the small instantons or ideal instantons. As an example, the Yang-Mills action density \eqref{YMaction} is actually singular in the $\lambda\rightarrow 0$ limit. In the noncommutative spaces, however, the small instanton singularities are resolved and the small instantons become smooth instantons 
which are new physical objects \cite{NeSc, Furuuchi}. Therefore, we can also expect that the singularity in the 4dWZW action densities \eqref{WZWaction} could be resolved in the noncommutative spaces and some new physical objects might  be found. Fortunately, our solutions are represented by quasideterminants which are straightforward tools to extend our solutions to noncommutative settings \cite{GHHN}. Furthermore, the discussion of noncommutative Ward conjecture and various examples are summarized e.g. in \cite{Hamanaka_NPB, Hamanaka_PLB, HaTo}. 
Some other developments in noncommutative integrable systems related to quasideterminants can be found in e.g. \cite{BRRS, CaSc, GLS,HaOk, Huang2, Li, Mikhailov, CLM, NiYi} as well.
In physical sense, it is no difficulty to introduce background $B$-fields \cite{Hull} into the N=2 string theory (cf. \cite{GOS,LPS}),  
therefore, noncommutative version of the unified theory (6dCS$\rightarrow$ 4dCS/4dWZW) of integrable systems could be proposed in the split signature. 

\begin{wrapfigure}[11]{r}{7.2cm}
 \centering
\vspace{-0.3cm}
\hspace{1.5cm}
\includegraphics[width=7.6cm]{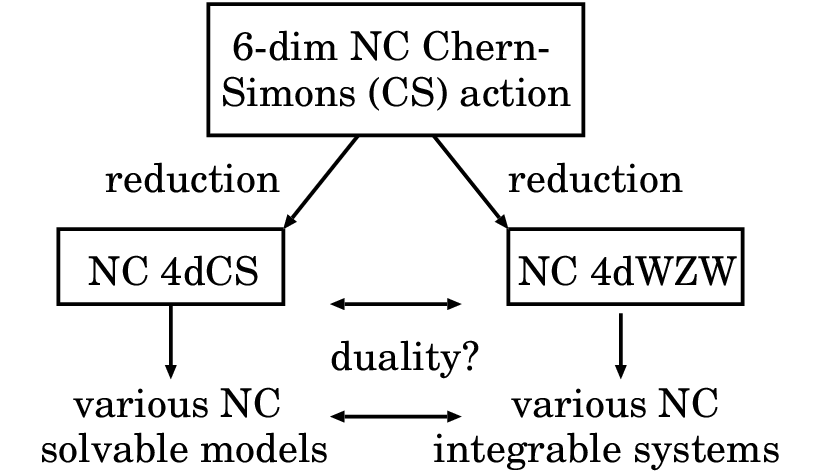}
\label{NCunify}
\end{wrapfigure}
Finally we would like to comment on quantization of integrable systems in homotopy algebra formulations. Lagrange formalisms of field theories can be reformulated in terms of homotopy algebras. This formulation has played important roles in  string field theories (for a review, see e.g. \cite{SeZw}) and can be applied to the open N=2 string theory as well. Recently Yuji Okawa gave a formula of correlation functions for wide class of field theories in this formalism \cite{Okawa}. This result can be applied to the open N=2 string field theory action to prove that $n(\geq 4)$-point functions vanish. On the other hand, there are other Lagrange formalisms for integrable systems in which infinite symmetry is manifestly considered: Lagrangian multi-form theories developed by Frank Nijhoff et al. (e.g. \cite{LoNi}) and pluri-Lagrangian systems developed by Yuri Suris et al. (e.g. \cite{BoSu}). These Lagrange formalisms of integrable systems could be formulated in terms of homotopy algebras as well. If our prediction is correct, Okawa's formula would show property of S-matrix factorizations, and as the result the Yang-Baxter equations in a universal way. 
This would be a bridge between classical integrability related to infinite symmetry and quantum integrability related to the Yang-Baxter equations. Of course there might be a possibility that integrability is not preserved in the process of quantizations. In this case, it could be interpreted as the ``integrability anomaly.'' 
In summary,  homotopy algebra formulations of the Lagrangian multi-form theory and the pluri-Lagrangian system would give a systematic quantization of integrable systems to clarify relationship between classical integrability and quantum integrability together with the integrability anomaly. 
Lagrangian multi-form formulation of (6dCS$\rightarrow$ 4dCS/4dWZW) is also a challenging problem. (cf. \cite{MNR})

\medskip
\noindent
{\bf Note added:} 
The authors thank Richard Szabo for informing us, after submission, that the connections between noncommutative deformations of ASDYM equations, open N=2 strings in background B-fields, and homotopy algebras has been elucidated in \cite{SzTr, Szabo}. This sugggests that our results and future directions would have interesting interpretations in terms of the double copy prescriptions as in \cite{SzTr, Szabo}. 

\subsection*{Acknowledgements}

The authors thank Hiroaki Kanno for collaboration at the early stage of
this work. MH thanks the organizers, especially Frank Nijhoff, for invitation, hospitality and discussion at the BIRD-IASM workshop on Lagrangian Multiform Theory and Pluri-Lagrangian Systems, Hangzhou, China in October 2023, and the organizers, especially Norbert Euler, for hospitality at  Open Communications in Nonlinear Mathematical Physics (OCNMP) workshop, Bad Ems, Germany in June 2024. MH is also grateful to string group members at Nagoya university for useful comments at the String Journal Club on July 24, 2023. Thanks are due to the YITP at Kyoto University, where he had fruitful discussions at the conference on Strings and Fields 2023 (YITP-W-23-07). The work of MH is supported in part by the Ichihara International Foundation and the Chubei Ito Foundation. The work of SCH is supported in part by Grant-in-Aid for Scientific Research (\#18K03274), 
the Iwanami Fujukai Foundation and the Daiko Foundation.

\label{lastpage}

\begin{thebibliography}{99}


\bibitem{Berkovits}
N.~Berkovits,
``SuperPoincare invariant superstring field theory,''
Nucl. Phys. B \textbf{450}, 90-102 (1995)
[erratum: Nucl. Phys. B \textbf{459}, 439-451 (1996)]
[arXiv:hep-th/9503099].


\bibitem{Bittleston}
R.~Bittleston, 
``Integrability from Chern-Simons theories,''
Ph.D thesis (2022, Cambridge U, DAMTP). 

\bibitem{BiSk}
R.~Bittleston and D.~Skinner,
``Twistors, the ASD Yang-Mills equations, and 4d Chern-Simons theory,''
JHEP \textbf{02}, 227 (2023) 
[arXiv:2011.04638].

\bibitem{BoSu}
A.~Bobenko, Yu.~B.~Suris, 
``On the Lagrangian structure of integrable quad-equations,'' 
Lett.\ Math.\ Phys.\ {\bf 92} (2010) 17–31. 

\bibitem{BRRS}
I.~Bobrova, V.~Retakh, V.~Rubtsov and G.~Sharygin, 
``Non-abelian discrete Toda chains and related lattices,''
Physica D {\bf 464}, 134200 (2024), [arXiv:2311.11124].

\bibitem{CaSc}
S.~Carillo and C~Schiebold, 
``Soliton Euations: admitted solutions and invariances via B\"acklund transformations,''
OCNMP, Special Issue in Memory of Decio Levi (2024) 12497. 

\bibitem{CDS}
A.~Connes, M.~R.~Douglas and A.~Schwarz,
``Noncommutative geometry and matrix theory: Compactification on tori,''
JHEP {\bf 9802} (1998) 003
{[hep-th/9711162]}.

\bibitem{Costello}
K.~Costello, 
``Topological strings, twistors, and Skyrmions,'' 
talk at Western Hemisphere Colloquium on Geometry and Physics (WHCGP)
on April 27, 2020. 

\bibitem{CoYa}
K.~Costello and M.~Yamazaki,
``Gauge Theory And Integrability, III,''
[arXiv:1908.02289].

\bibitem{DLMV}
F.~Delduc, S.~Lacroix, M.~Magro and B.~Vicedo,
``A unifying 2D action for integrable $\sigma $-models from 4D Chern-Simons theory,''
Lett. Math. Phys. \textbf{110}, 1645 (2020)
[arXiv:1909.13824].

\bibitem{deVega}
  H.~J.~de Vega,
  ``Nonlinear Multiplane Wave Solutions of Selfdual {Yang-Mills} Theory,''
  Commun.\ Math.\ Phys.\  {\bf 116}, 659 (1988).

\bibitem{Donaldson}
S.~K.~Donaldson, 
``Anti-self-dual Yang-Mills connections over complex algebraic surfaces and stable vector bundles,''
Proc.\ Lond.\ Math.\ Soc.\ {\bf 3} (1985) 1.

\bibitem{DoHu}
M.~R.~Douglas and C.~Hull,
``D-branes and the noncommutative torus,''
JHEP {\bf 9802} (1998) 008
{ [hep-th/9711165]}.

\bibitem{FSY}
O.~Fukushima, J.~Sakamoto and K.~Yoshida,
``Non-Abelian Toda field theories from a 4D Chern-Simons theory,''
JHEP \textbf{03}, 158 (2022)[arXiv:2112.11276].

\bibitem{Furuuchi}
K.~Furuuchi,
``Instantons on noncommutative $\mathbb{R}^4$ and projection operators,''
Prog. Theor. Phys. \textbf{103}, 1043-1068 (2000)
[arXiv:hep-th/9912047].

\bibitem{GWW}
  M.~L.~Ge, L.~Wang and Y.~S.~Wu,
  ``Canonical reduction of selfdual Yang-Mills theory to sine-Gordon and Liouville theories,''
  Phys.\ Lett.\ B {\bf 335} (1994) 136.

\bibitem{GGRW}
I.~Gelfand, S.~Gelfand, V.~Retakh and R.~Wilson,
``Quasideterminants,''
Adv.\ Math.\ {\bf 193} (2005) 56
[math.QA/0208146].

\bibitem{GeRe}
I.~Gelfand and V.~Retakh,
``Determinants of matrices over noncommutative rings,''
Funct.\ Anal.\ Appl.\ {\bf 25} (1991) 91; 
``A theory of noncommutative determinants and characteristic functions of graphs,''
Funct.\ Anal.\ Appl.\ {\bf 26} (1992) 231.

\bibitem{GHHN}
C.~R.~Gilson, M.~Hamanaka, S.C.~Huang and J.~J.~C.~Nimmo, 
``Soliton solutions of noncommutative anti-self-dual Yang-Mills equations,''
J. Phys. A \textbf{53}, 404002 (2020)
[arXiv:2004.01718].

\bibitem{GLS}
C.~Gilson, S.H.~Li and Y.~Shi
``Matrix-valued $\theta$-deformed bi-orthogonal polynomials, Non-commutative Toda theory and B\"acklund transformation,''
[arXiv:2305.17962]. 

\bibitem{GOS}
D.~Gluck, Y.~Oz and T.~Sakai,
``D-branes in N=2 strings,''
JHEP \textbf{08}, 055 (2003)
[arXiv:hep-th/0306112].

\bibitem{Hamanaka_NPB}
  M.~Hamanaka,
  ``Noncommutative Ward's conjecture and integrable systems,''
  Nucl.\ Phys.\  B {\bf 741} (2006) 368
  [hep-th/0601209].

\bibitem{Hamanaka_PLB}
M.~Hamanaka,
``On reductions of noncommutative anti-self-dual Yang-Mills equations,''
Phys. Lett. B \textbf{625}, 324-332 (2005)
[hep-th/0507112]. 

\bibitem{HaOk} 
  M.~Hamanaka and H.~Okabe,
  ``Soliton Scattering in Noncommutative Spaces,''
  Theor.\ Math.\ Phys.\  {\bf 197}, 1451 (2018)
  [Teor.\ Mat.\ Fiz.\  {\bf 197}, no. 1, 68 (2018)]
  [arXiv:1806.05188].

\bibitem{HaHu}  
M.~Hamanaka and S.C.~Huang,
``New Soliton Solutions of Anti-Self-Dual Yang-Mills Equations,''
JHEP \textbf{10}, 101 (2020) 
[arXiv:2004.09248].


\bibitem{HaHu2}
M.~Hamanaka and S.C.~Huang,
``Multi-soliton dynamics of anti-self-dual gauge fields,''
JHEP \textbf{01}, 039 (2022)
[arXiv:2106.01353].

\bibitem{HHK}
M. Hamanaka, S.C. Huang and H. Kanno, 
``Solitons in Open N=2 String Theory,''\\ 
PTEP \textbf{2023}, no.4, 043B03 (2023) 
{[arXiv: 2212.11800]}.

\bibitem{HaTo}
 M.~Hamanaka and K.~Toda,
 ``Towards noncommutative integrable systems,''
 Phys.\ Lett.\ A {\bf 316}, 77 (2003) [hep-th/0211148].

\bibitem{Huang}  
S.~C.~Huang,
``On Soliton Solutions of the Anti-Self-Dual Yang-Mills Equations from the Perspective of Integrable Systems,''
Ph.D thesis (2021, Nagoya University)
[arXiv:2112.10702].

\bibitem{Huang2}
S.~C.~Huang,
``Multi-Soliton scattering of the Anti-Self-Dual Yang-Mills Equations in 4-dimensional split signature,'' 
in Proceedings of the East Asia Joint Symposium on Fields and Strings 2021, pp. 33-42 (2022), 
[arXiv:2201.13318].

\bibitem{Hull}
C.~M.~Hull,
``The Geometry of N=2 strings with torsion,''
Phys. Lett. B \textbf{387}, 497-501 (1996)
[hep-th/9606190].

\bibitem{IKUX}
T.~Inami, H.~Kanno, T.~Ueno and C.~S.~Xiong,
``Two toroidal Lie algebra as current algebra of four-dimensional Kahler WZW model,''
Phys. Lett. B \textbf{399}, 97-104 (1997)
[arXiv:hep-th/9610187].

\bibitem{Kodama}
Y.~Kodama,
{\it KP Solitons and the Grassmannians},
(Springer, 2017).


\bibitem{KoWi}
Y.~Kodama and L.~Williams,
``KP solitons and total positivity for the Grassmannian,''
Invent.\  math.\  {\bf 198}, 637 (2014) 
[arXiv:1106.0023].

\bibitem{KMO}
D.~Kutasov, E.~J.~Martinec and M.~O'Loughlin,
``Vacua of M theory and N=2 strings,''
Nucl. Phys. B \textbf{477}, 675-700 (1996)
[arXiv:hep-th/9603116].

\bibitem{LPS}
O.~Lechtenfeld, A.~D.~Popov and and B.~Spendig,
``Noncommutative solitons in open N= 2 string theory,''
JHEP {\bf 06} (2001) 011
[hep-th/0103196].

\bibitem{Lee}
K.~M.~Lee,
``Sheets of BPS monopoles and instantons with arbitrary simple gauge group,''
Phys. Lett. B \textbf{445}, 387-393 (1999)
[arXiv:hep-th/9810110].

\bibitem{Li}
C.~Li, 
``Symmetries and Reductions on the noncommutative Kadomtsev-Petviashvili and Gelfand-Dickey hierarchies,''
J. Math. Phys. 59, 123503 (2018),  
[arXiv:1907.04169]. 

\bibitem{LiZh} 
S.S.~Li and D.J.~Zhang, 
``Direct linearization of the SU(2) anti-self-dual Yang-Mills equation in various spaces,''
[arXiv:2403.06055].

\bibitem{LQZ} 
S.S.~Li, C.Z.~Qu and D.J.~Zhang, 
``Solutions to the SU(N) self-dual Yang-Mills equation,''
Physica D {\bf 453} (2023) 133828 
[arXiv:2211.08574].

\bibitem{LQYZ} 
S.S.~Li, C.Z.~Qu, X.X.~Yi  and D.J.~Zhang, 
``Cauchy matrix approach to the SU(2) self-dual Yang--Mills equation,'' 
Stud.\ Appl.\ Math.\ {\bf 148} (2022) 1703
[arXiv:2112.06408]. 

\bibitem{LoNi}
S.~Lobb and F.~W.~Nijhoff, 
``Lagrangian multiforms and multidimensional consistency,'' 
J.\ Phys.\ A {\bf 42} (2009) 454013. 
[arXiv:0903.4086]

\bibitem{LMNS}
A.~Losev, G.~W.~Moore, N.~Nekrasov and S.~Shatashvili,
``Four-dimensional avatars of two-dimensional RCFT,''
Nucl. Phys. B Proc. Suppl. \textbf{46}, 130-145 (1996)
[arXiv:hep-th/9509151].

\bibitem{Marcus}
N.~Marcus,
``The N=2 Open String,''
Nucl.\ Phys.\ B {\bf 387} (1992) 263
[hep-th/9207024]; 
``A tour through N=2 strings,''
hep-th/9211059.

\bibitem{MNR}
J.~F.~Martins, F.~W.~Nijhoff and D.~Riccombeni,
``Darboux-Kadomtsev-Petviashvili system as an integrable Chern-Simons multiform theory in infinite dimensional space,''
Phys. Rev. D \textbf{109}, L021701 (2024)
[arXiv:2305.03182].
	
\bibitem{Mason}
L.~J.~Mason,
``Global anti-self-dual Yang-Mills fields in Ultrahyperbolic signature and their scattering,''
J.\ reine angew.\ Math.\ {\bf 597}, 105 (2007) 
[math-ph/0505039].

\bibitem{MaSp}
L.~J.~Mason and G.~A.~J.~Sparling,
``Nonlinear Schrodinger and Korteweg-de Vries Are Reductions of Selfdual Yang-Mills,''
Phys. Lett. A \textbf{137}, 29-33 (1989)

\bibitem{MaWo}
L.~J.~Mason and N.~M.~Woodhouse,
{\it Integrability, Self-Duality, and Twistor Theory}
(Oxford UP, 1996).

\bibitem{Mikhailov} 
A.~V.~Mikhailov, 
``Quantisation ideals of nonabelian integrable systems,'' 
Russian Math.\ Surveys {\bf 75} (2020) 978-980. 

\bibitem{CLM}
K.~Muhammad, C.~X.~Li and M.~Cui, 
``The extended versions of the noncommutative KP and mKP equations and Miura transformation,'' 
[arXiv:2404.11391]. 

\bibitem{NaSc}
V.~P.~Nair and J.~Schiff,
``A Kahler-{Chern-Simons} Theory and Quantization of Instanton Moduli Spaces,''
Phys. Lett. B \textbf{246}, 423-429 (1990);

\bibitem{NaSc2}
V.~P.~Nair and J.~Schiff,
``Kahler Chern-Simons theory and symmetries of antiselfdual gauge fields,''
Nucl. Phys. B \textbf{371}, 329-352 (1992).

\bibitem{Nakajima}
H.~Nakajima, 
``Resolutions of moduli spaces of ideal instantons on $\mathbb{R}^4$,''
in {\it Topology, Geometry and Field Theory}
(World Sci., 1994) 129 {[ISBN/981-02-1817-6]}.

\bibitem{NeVa}
A.~Neitzke and C.~Vafa,
``N=2 strings and the twistorial Calabi-Yau,''
[arXiv:hep-th/0402128].

\bibitem{Nekrasov}
N.~Nekrasov,
``Four Dimensional Holomorphic Theories,''
Ph. D thesis (1996, Princeton U).

\bibitem{NeSc}
N.~Nekrasov and A.~Schwarz, 
``Instantons on noncommutative $\mathbb{R}^4$, and (2,0) superconformal six dimensional theory,''
Commun.\ Math.\ Phys.\ {\bf 198} (1998) 689
{ [hep-th/9802068]}.

\bibitem{GNO} 
  J.~J.~C.~Nimmo, C.~R.~Gilson and Y.~Ohta,
  ``Applications of Darboux transformations to the selfdual Yang-Mills equations,''
  Theor.\ Math.\ Phys.\  {\bf 122}, 239 (2000)
  [Teor.\ Mat.\ Fiz.\  {\bf 122}, 284 (2000)].

\bibitem{NiYi}
J.~J.~C.~Nimmo and H.~Yilmaz, 
``Binary Darboux transformation for the Sasa-Satsuma equation,''
J.\ Phys.\ A {\bf 48}, 425202 (2015)
[arXiv:1502.07371].

\bibitem{OhWa}
K.~Ohkuma and M.~Wadati, 
``The Kadomtsev-Petviashvili Equation: the Trace Method and the Soliton Resonances,''
J.\ Phys.\ Soc.\ Jap.\  {\bf 52}, 749 (1983).

\bibitem{Ohta}
Y.~Ohta, ``Recent topics on rogue waves,''
talk at RIMS workshop on Recent Developments of Integrable Systems, Kyoto, September 2023.

\bibitem{Okawa}
Y.~Okawa,
``Correlation functions of scalar field theories from homotopy algebras,''
JHEP \textbf{05}, 040 (2024)
[arXiv:2203.05366].

\bibitem{OoVa}
H.~Ooguri and C.~Vafa,
``N=2 heterotic strings,''
Nucl.\ Phys.\ B {\bf 367} (1991) 83.
	
\bibitem{Parkes}
A.~Parkes,
``On N=2 strings and classical scattering solutions of selfdual Yang-Mills in (2,2) space-time,''
Nucl. Phys. B \textbf{376}, 279-296 (1992)
[arXiv:hep-th/9110075].

\bibitem{SeWi}
N.~Seiberg and E.~Witten,
``String theory and noncommutative geometry,''
JHEP \textbf{09}, 032 (1999)
[arXiv:hep-th/9908142].

\bibitem{SeZw}
A.~Sen and B.~Zwiebach,
``String Field Theory: A Review,''
[arXiv:2405.19421].

\bibitem{Szabo}
R.~J.~Szabo,
``Gravity versus Noncommutative Gauge Theory: A Double Copy Perspective,''
[arXiv:2401.16283].

\bibitem{SzTr}
R.~J.~Szabo and G.~Trojani,
``Homotopy Double Copy of Noncommutative Gauge Theories,''
Symmetry \textbf{15}, no.8, 1543 (2023)
[arXiv:2306.12175].

\bibitem{Ward}
R.~S.~Ward,
``Integrable and solvable systems, and relations among them,''
Phil.\ Trans.\ Roy.\ Soc.\ Lond.\ A {\bf 315} (1985) 451.

\bibitem{Witten}
E.~Witten,
``Some Exact Multi - Instanton Solutions of Classical Yang-Mills Theory,''
Phys. Rev. Lett. \textbf{38}, 121-124 (1977). 

\bibitem{Yang}
C.~N.~Yang,
``Condition of Selfduality for SU(2) Gauge Fields on Euclidean Four-Dimensional Space,'' 
Phys. Rev. Lett. \textbf{38}, 1377 (1977). 

\end{thebibliography}
\end{document}